%% file: GRMM.tex
\documentclass[sigconf,nonacm]{acmart}

\usepackage{enumitem}
\usepackage{multicol}
\usepackage{soul}
\usepackage{hyperref}       
\usepackage{url}            
\usepackage{booktabs}       
\usepackage{array}          
\usepackage{xcolor}         
\usepackage{colortbl} 
\usepackage{cuted,tcolorbox}

\newtheorem{definition}{Definition}

\AtBeginDocument{%
  }

\acmConference[SIGIR '26]{Make sure to enter the correct
  conference title from your rights confirmation email}{July 03--05,
  2026}{Melbourne, Australia}
\acmISBN{978-1-4503-XXXX-X/2018/06}

\newcommand{\frameworkname}{\textit{MMGRid }}

\begin{document}

\title{MMGRid: Navigating Temporal-aware and Cross-domain Generative Recommendation via Model Merging}

\input{Main/authors}

\renewcommand{\shortauthors}{Wei et al.}

\begin{abstract}
\textbf{Model merging (MM)} offers an efficient mechanism for integrating multiple specialized models without access to original training data or costly retraining. While MM has demonstrated success in natural language processing and computer vision, its role in recommender systems (RSs) remains largely unexplored.
Recently, \textbf{Generative Recommendation (GR)} has emerged as a new paradigm in RSs, characterized by rapidly growing model scales and substantial computational costs, making MM particularly appealing for cost-sensitive deployment scenarios.
In this work, we present the first systematic study of MM in GR through a contextual lens. We focus on a fundamental yet underexplored challenge in real-world GR: \textit{how to merge generative recommenders specialized to different real-world contexts, arising from temporal evolving user behaviors and heterogeneous application domains.}
To this end, we propose a unified framework \frameworkname, a structured contextual \textbf{\underline{grid}} of GR checkpoints that organizes models trained under diverse contexts induced by \textbf{temporal evolution and domain diversity}. All checkpoints are derived from a shared base large language model (LLM) but fine-tuned on context-specific data, forming a realistic and controlled model space for systematically analyzing MM across GR paradigms and merging algorithms.
Our investigation reveals several key insights. \textbf{First, training GR models from LLMs can introduce parameter conflicts during merging due to token distribution shifts and objective disparities}; such conflicts can be alleviated by disentangling task-aware and context-specific parameter changes via base model replacement. \textbf{Second, incremental training across contexts induces recency bias, which can be effectively balanced through weighted contextual merging}. Notably, we observe that optimal merging weights correlate with context-dependent interaction characteristics, offering practical guidance for weight selection in real-world deployments.
We open-source our framework, training pipelines, evaluation protocols, and contextual checkpoints to support future research on MM in GR. 
\end{abstract}




\maketitle

\section{Introduction}
In machine learning, \textbf{Model merging (MM)}~\citep{yangModelMerging2024} has recently emerged as an efficient paradigm for integrating the capabilities of multiple specialized models into a single, universal model without requiring access to the original training data or additional expensive optimization.
Building on the idea that fine-tuned models can be represented as parameter deltas from a shared pretrained backbone, MM directly operates on model parameters, typically through linear interpolation~\citep{wortsmanModelSoups2022}, task-vector arithmetic~\citep{ilharcoEditingModels2022}, or geometry-aware fusion~\citep{yadavTIESMergingResolving2023}, to aggregate diverse features, skills, or domain knowledges. 
This property naturally aligns with modern machine learning pipelines, which follow the ubiquitous "pre-training → post-training" paradigm~\citep{kumarLLMPostTraining2025, zhuangRobustlyOptimized2021}. Moreover, under the influence of scaling laws~\citep{kaplanScalingLaws2020}, increasing model sizes further exacerbate the cost of retraining from scratch. MM bypasses this costly process by directly merging post-trained models, offering a more efficient and scalable alternative to traditional training pipelines.

The effectiveness of MM in reusing and recombining specialized expertise has led to its widespread adoption in computer vision~\citep{ilharcoEditingModels2022,wortsmanModelSoups2022}, natural language processing (NLP)~\citep{akibaEvolutionaryOptimization2025}, and multimodal learning~\citep{aielloJointlyTraining2023}.
\textbf{Recommendation Systems (RSs)}, as an application-driven machine learning problem, are also encompassed within this broad paradigm.
However, traditional RSs, characterized by multi-stage pipelines of heterogeneous components built on domain-specialized signals, have offered little incentive for parameter-level fusion~\citep{dengOneRecUnifying2025}.
This situation is fundamentally changing with the recent emergence of \textbf{Generative Recommendation (GR)}~\citep{rajputRecommenderSystems2023a,tanPCRCAParallel2025,chenHLLMEnhancing2024,zhaiActionsSpeak2024,duReinforcementSpeculative2025a}. Inspired by the evolution of NLP towards Large Language Models (LLMs), GR replaces fragmented multi-stage architectures with end-to-end autoregressive modeling~\citep{dengOneRecUnifying2025}, unifying representation learning, semantic reasoning, and decision making within a single generative framework.
This architectural unification not only enhances semantic expressiveness and cross-scenario generalization, but also introduces a training and deployment pattern closely aligned with LLMs, making MM a technically viable yet previously unexplored strategy for GR.
Furthermore, real-world RS deployments face persistent constraints, including stringent cost sensitivity~\citep{zhouOneRecV2Technical2025} and the need for continual model adaptation driven by ever-evolving user behaviors and streaming interaction data~\citep{guoEnhancingNewitem2025,yooContinualRecommender2025}. MM, with its ability to efficiently recombine specialized checkpoints, is uniquely positioned to address such contextual variation.
However, despite this strong practical motivation, a systematic understanding of how MM behaves under realistic contextual shifts in GR remains missing.

\begin{figure*}[!t]
    \centering
    \includegraphics[width=0.82\linewidth]{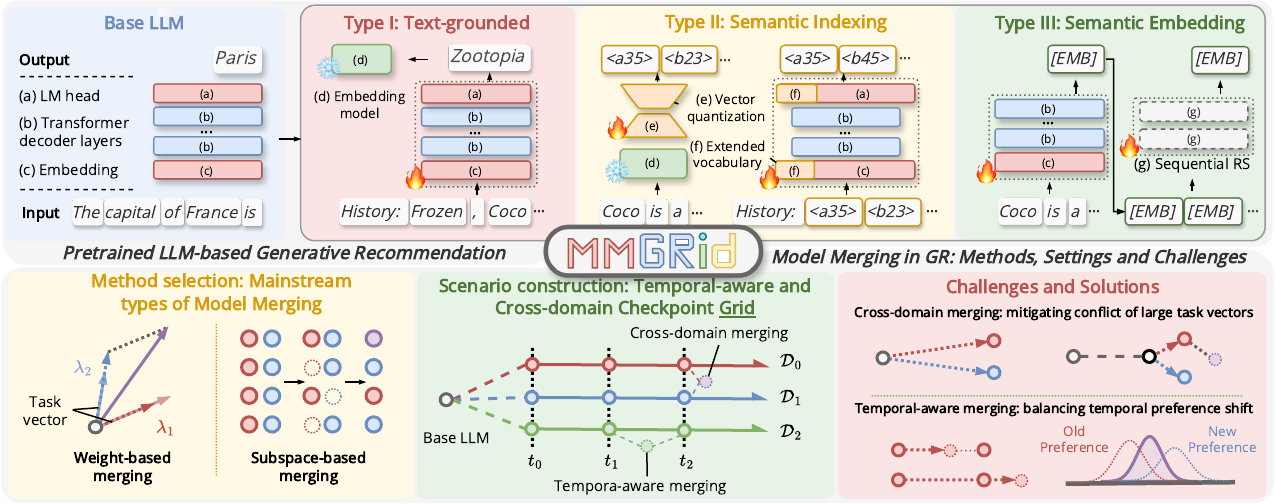}
    \caption{Overview of the \frameworkname Framework. Generative recommendation (GR) models built on a shared base LLM are organized into a contextual grid across domains and temporal stages. This grid provides a unified model space for studying merging of GR models on contextual scenarios, identifying key challenges and promising directions for solutions.}
    \Description{
        Overview of the \frameworkname Framework. Generative recommendation (GR) models built on a shared base LLM are organized into a contextual grid across domains and temporal stages. This grid provides a unified model space for studying merging of GR models on contextual scenarios, identifying key challenges and promising directions for solutions.
    }
    \label{fig:Overview}
\end{figure*}

In this work, we present the first systematic investigation of MM for GR from a \emph{contextual perspective}.
We introduce a unified experimental framework, \frameworkname, which constructs a \emph{contextual grid} as a structured model space composed of GR checkpoints specialized to different real-world contexts.
The framework encompasses three representative GR paradigms: text-grounded generation~\citep{baoBiStepGrounding2025}, semantic ID modeling~\citep{dengOneRecUnifying2025,zhengAdaptingLarge2024}, and semantic embedding modeling~\citep{chenHLLMEnhancing2024}, and two mainstream merging algorithms: weight-based merging~\citep{ilharcoEditingModels2022} and subspace-based merging~\citep{yadavTIESMergingResolving2023}, together with consistent training and inference pipelines.
Building upon this contextual model space, we study MM along two fundamental dimensions of real-world context: \textbf{heterogeneous recommendation domains} and \textbf{temporal evolution of user interactions}.
Rather than treating these as isolated settings, we view them as complementary sources of contextual variation that jointly shape the deployment landscape of GR.
This perspective enables a unified analysis of merging behaviors across contexts that naturally arise in cross-domain recommendation and continual system updating.

Through extensive experiments on the contextual checkpoint grid built on six diverse domains, we identify several key challenges and derive actionable insights:
\begin{itemize}[leftmargin=12pt]
    \item \textbf{Creating unifed GR model via cross-domain merging:} LLM-based GR models may exhibit large task-aware parameter shifts due to differences in data distribution and learning objectives. These shifts become entangled with domain-specific knowledge, leading to severe conflicts and performance degradation in cross-domain merging. By leveraging checkpoint information from incremental training, we demonstrate the potential to decouple these two types of parameter changes, thereby improving cross-domain merging performance.
    
    \item \textbf{Capturing evolving user preference via temporal-aware merging:} In incremental training for RSs, parameter shifts induced by newly added data pose challenges in balancing users' evolving and historical preferences. We empirically confirm that weighted fusion of checkpoints from different temporal snapshots can better balance the impact of temporal preference shifts. Furthermore, we reveal the connection between cross-domain temporal preference transfer patterns and merging weights, enabling efficient achievement of overall optimal performance through model merging.
\end{itemize}
These explorations provide important insights into the applicability and challenges of MM for GR, revealing both the opportunities and limitations of parameter-level merging in recommendation scenarios. We believe that our findings will inform the design of more effective merging strategies tailored to the unique characteristics of GR models. Additionally, to promote reproducibility and encourage further investigation, we open-source our experimental framework \footnote{\url{https://github.com/Joinn99/MMGRid}}, trained models, and evaluation protocols, which we hope will serve as a foundation for future research in this emerging direction.

\section{Preliminaries}
\subsection{Generative Recommendation}
\label{ssec:gr}
In this paper, we focus on GR within the context of \textbf{Sequential Recommendation (SeqRec)}~\citep{adomaviciusNextGeneration2005,kangSelfAttentiveSequential2018,sunLLM4RSRLarge2025,liuDiagnosticGuidedDynamic2026}, which is a fundamental and representative scenario in RSs, aiming to model the users' dynamic interest via their sequential interaction history.
Formally, a SeqRec task is defined as follows: 
\begin{definition}[Sequential Recommendation]
Let $\mathcal{U}$ be the set of users and $\mathcal{I}$ be the set of items. For a user $u \in \mathcal{U}$, and the historical interaction sequence of $u$ is $S_u = (i_1, i_2, \dots, i_T)$, where $i_k \in \mathcal{I}$ is the $k$-th item $u$ interacted with, and $T$ is the sequence length. The objective is to predict the item $i_{T+1}$ that the user $u$ is most likely to interact with in the future.
\end{definition}

GR is an emerging paradigm that applies generative models, such as auto-regressive Transformer decoders successful in NLP to RSs. This approach reframes the recommendation task as a sequence generation problem, where the model learns to generate a sequence of recommended items (or related text) conditioned on the user's history and context.

A core challenge in adapting generative models to recommendation is determining the input and output representation for items, which serves as the fundamental element of a RS. GR aims to leverage the powerful common-sense reasoning and generalization capabilities demonstrated by LLMs in NLP~\citep{yangQwen3Technical2025}. To achieve this, they seek to learn from both the semantic information inherent in items (e.g., text descriptions, images) and the collaborative information embedded in user interaction patterns~\citep{baoBiStepGrounding2025,zhangDualPhasePlaytimeguided2025}.

In literature, various approaches have been proposed to address this item representation challenge. As shown in Figure \ref{fig:Overview}, we categorize the existing paradigms into three main types based on the representation space of items in the generative models:
\begin{itemize}[leftmargin=12pt]
    \item \textbf{Text-Grounded:} These methods use the raw semantic information of an item, such as its title or detailed description, as the direct input and output from the generative model~\citep{baoBiStepGrounding2025, baoDecodingMatters2024, dengLogitSpace2025, dholeGenerativeProduct2025}. The model learns to generate item titles or other textual identifiers directly as the recommendations, which are then mapped to the corresponding items in the item space. 

    \item \textbf{Semantic Indexing (SemID):} To avoid the hallucination issue in text-grounded generation, this approach uses techniques like Vector Quantization (VQ)~\citep{vandenoordNeuralDiscrete2017} and clustering to encode rich semantic information into a finite set of discrete indices or "semantic IDs". These new IDs are then treated as new tokens in the generative model's vocabulary, effectively tokenizing the item space based on semantic meaning rather than arbitrary indices~\citep{dengOneRecUnifying2025, rajputRecommenderSystems2023a, zhengAdaptingLarge2024}.

    \item \textbf{Semantic Embedding (SemEmb):} Similar to traditional sequential recommendation, these methods first encode an item's semantic information into a single, dense embedding. This embedding is then used as the input and output representation of items for the generative model~\citep{chenHLLMEnhancing2024, heLLM2RecLarge2025, linOrderagnosticIdentifier2025}. 
\end{itemize}
It is noteworthy that some recent methods employ a hybrid approach, combining two or more of these representations to capture a richer and more robust signal~\citep{yangSparseMeets2025,yangUnifyingGenerative2025}. Although these varied methodologies differ substantially, they are uniformly characterized by the strategic utilization of a pretrained base model, which incorporates extensive world knowledge, to significantly enhance the efficacy of GR.

\subsection{Model Merging}
Model Merging (MM) is an effective technique that merges the parameters of multiple distinct models to create a single unified model with various objectives.
A significant advantage of MM is its efficiency: it bypasses the substantial computational expense and the need for original training data associated with retraining from scratch~\citep{wortsmanModelSoups2022,ilharcoEditingModels2022}. The merging is performed directly at the parameter level, resulting in a single model for inference. 
Here, we define the parameter-wise model merging problem as follows:
\begin{definition}[Parameter-wise Model Merging]
Assume we have $N$ models, $\Phi_{\Theta^{(1)}}, \dots, \Phi_{\Theta^{(N)}}$, which share the same architecture. These models are typically fine-tuned from a common pre-trained model $\Phi_{\Theta}$. The parameters (or weights) of the $n$-th model are denoted as $\Theta^{(n)}$. 
The objective of model merging is to find a function $\textsc{merge}(\cdot)$ that combines the parameters of these models to produce a new set of parameters $\Theta^{*}$:
\begin{equation}
\Theta^{*} = \textsc{Merge}(\Theta^{(1)}, \dots, \Theta^{(N)}).
\end{equation}
\end{definition}

Recently, the emergence of the pretrained–finetuning paradigm linked the parameter changes to the specific downstream task, providing a clear connection between parameter adaptations and task identity.
The concept of the \textbf{task vector} was then introduced, providing a structured way to represent these connections~\citep{ilharcoEditingModels2022}. 
A task vector $\boldsymbol{\tau}_k$ for a task $k$ is defined as the difference between the fine-tuned model parameters $\Theta^{(k)}$ and the pre-trained base model parameters $\Theta^{(0)}$:
\begin{equation}
\boldsymbol{\tau}(\Theta^{(0)}, \Theta^{(k)}) = \Theta^{(k)} - \Theta^{(0)}.
\end{equation}
Task vectors are understood to capture the essential modifications needed for a model to perform a particular task, thereby guiding the base model's behavior. Notably, representing parameters as task vectors permits compositional operations: for example, combining the capabilities of multiple tasks via addition, or reducing a specific task's influence via subtraction~\citep{ilharcoEditingModels2022}.

Given the general formulation above, a variety of concrete task-vector MM methods for the function $\textsc{merge}(\cdot)$ have been proposed, which can be broadly classified into \textbf{weight-based} and \textbf{subspace-based} approaches.

\textbf{Weight-based merging.} This family of methods directly merges models by operating on their parameters~\citep{yangModelMerging2024,ilharcoEditingModels2022}. The most common weight-based technique is the \emph{linear interpolation} of parameters, where the merged model's weights are defined as a convex (or more generally affine) combination of the input models:
\begin{equation}
\Theta^{*} = \sum_{n=1}^N \lambda_n \boldsymbol{\tau}(\Theta^{(0)}, \Theta^{(n)}),
\end{equation}
where $\lambda_n$ is the weight for the $n$-th model, which is typically set to be non-negative.

\textbf{Subspace-based merging.} Inspired by model pruning, several methods attempt to mitigate the merge conflicts by identifying and removing insignificant or redundant parameters in task vectors~\citep{yadavTIESMergingResolving2023}. The trimmed sparse task vectors are then used to construct the merged model:
\begin{equation}
\Theta^{*} = \frac{1}{N} \sum_{n=1}^N \textsc{Trim}(\boldsymbol{\tau}(\Theta^{(0)}, \Theta^{(n)})),
\end{equation}
where $\textsc{Trim}(\cdot)$ is the trimming function that removes insignificant or redundant parameters from the task vector. In this paper, we adopt two simple yet effective trimming methods simultaneously:
\begin{itemize}[leftmargin=12pt]
    \item \textbf{TIES~\citep{yadavTIESMergingResolving2023}:} Keep the top $X\%$ percantage of parameters with the largest magnitude in the task vector, and further trim the parameters with sign conflict by keeping the predomint sign across all task vectors.
    \item \textbf{DARE~\citep{yuLanguageModels2024}:} Randomly drop parameters with ratio $p$ in the task vector, and rescale the remaining parameters by a factor of $\frac{1}{1-p}$.
\end{itemize}

\section{Framework Design}
\label{sec:setup}
In GR, a significant characteristic is the manifestation of \textbf{scaling laws}~\citep{dengOneRecUnifying2025}.
The improved generalization capacity of larger-scale models allows them to effectively integrate knowledge from various domains. However, this increase in model size concurrently elevates the cost of model tuning for domain merging and incremental updates, posing a significant challenge to cost-sensitive recommendation scenarios with real-world contexts like diverse application domains and temporal evolving of interactions.
To address these challenges, MM techniques offer a promising solution by efficiently integrating the capabilities of multiple models through parameter space fusion without requiring additional training. In this work, we seek to create a experimantal framework, named \frameworkname, to explore the potential of MM in GR, identifying hidden issues and challenges and promising directions for solutions.

\subsection{Identifying Real-world Contexts of GR}
In this work, we identify two contexts in practical application that align closely with the characteristics and development trends of RSs, serving as the settings for model merging exploration:

\textbf{Cross-domain merging.} Following a development trajectory similar to NLP, GR are exhibiting a trend toward integrating different domains to construct unified models~\citep{zhouRecBaseGenerative2025}.
Traditional RSs are typically trained on domain-specific data, and may need to be merged to establish more unified models.
This approach enables leveraging multi-domain knowledge to enhance model capabilities while reducing deployment costs and avoiding the overhead of retraining models from scratch.
It is noted that this scenario is different from the traditional cross-domain RS~\citep{khanCrossDomain2017} which involves domain knowledge transfer in model training stage. In this scenario, all domain models are trained independently and the merging is totally under parameter level without additional training data involved.
\begin{definition}[Cross-Domain Merging]
    \label{def:cross_domain_merging}
    Suppose we have models $\Phi_{\Theta^{(\mathcal{D}_1, t_1)}}, \dots, \Phi_{\Theta^{(\mathcal{D}_N, t_1)}}$, which are trained on different $N$ domains $\mathcal{D}_1, \dots, \mathcal{D}_N$ at time $t_1$ respectively. The objective of cross-domain merging is to find a function $\textsc{Merge}(\cdot)$ that combines the parameters of these models to produce a new set of parameters $\Theta^{*}$:
   \begin{equation}
    \Theta^{*} = \textsc{Merge}(\Theta^{(\mathcal{D}_1, t_1)}, \dots, \Theta^{(\mathcal{D}_N, t_1)}).
   \end{equation}
\end{definition}

\textbf{Temporal merging.} In production environment, most RSs operate under an incremental training paradigm, continuously updating models based on streaming online data~\citep{guoEnhancingNewitem2025, yooContinualRecommender2025}. Across different time checkpoints, newer models trained on fresh data may capture evolving trends in overall user preferences. These temporal dynamics offer valuable insights that could enhance the overall performance of RSs. By merging models from different time periods, we can potentially synthesize both stable long-term patterns and emerging short-term trends, creating a more robust and adaptive recommendation model that balances historical knowledge with recent behavioral shifts.
\begin{definition}[Temporal Merging]
    Suppose we have models $\Phi_{\Theta^{(\mathcal{D}, t_1)}}, \dots, \Phi_{\Theta^{(\mathcal{D}, t_N)}}$, which are trained on the same domain $\mathcal{D}$ at time $t_1, \dots, t_N$ respectively. The objective of temporal merging is to find a function $\textsc{merge}(\cdot)$ that combines the parameters of these models to produce a new set of parameters $\Theta^{*}$:
    \begin{equation}
        \Theta^{*} = \textsc{Merge}(\Theta^{(\mathcal{D}, t_1)}, \dots, \Theta^{(\mathcal{D}, t_N)}).
    \end{equation}
\end{definition}

These two scenarios represent fundamental challenges in RSs: spatial generalization across domains and temporal adaptation to evolving user behaviors, making them ideal testbeds for investigating the efficacy of model merging techniques in GR.

\subsection{Creating Contextual Checkpoint Grid}
In the context of GR, several factors introduce key distinctions from typical application domains of MM. The temporal nature of user interaction data, parametric variations across different types of GR approaches, and differences in backbone model architectures collectively pose new challenges for model merging in RSs. In this work, we aim to closely approximate realistic RS scenarios and their practical requirements for model merging. Therefore, we construct a unified experimental framework that enables the creation of different types of GR models across \textbf{various domains and time periods}, all built upon \textbf{a unified backbone}, facilitating MM experiments under conditions that closely mirror real-world scenarios and align with practical demands.

\begin{table}[t!]
    \centering
    \renewcommand{\arraystretch}{0.8}
    \scriptsize
    \setlength\tabcolsep{2pt}
    \setul{1pt}{0.4pt}
    \caption{Statistics of the interaction data in Amazon dataset.}
    \label{tab:data_split}
    \begin{tabular}{@{}lrrrrrr@{}}
      \toprule
      \textbf{Domain} & \multicolumn{1}{c}{\textbf{\# Users}} & \multicolumn{1}{c}{\textbf{\# Items}} & \multicolumn{1}{c}{\textbf{\# Pretrain}} & \multicolumn{1}{c}{\textbf{\# $P_1$}} & \multicolumn{1}{c}{\textbf{\# $P_2$}} & \multicolumn{1}{c}{\textbf{\# Test}} \\ \midrule
      Video Games (Vid.) & 4,876 & 4,068 & 215,761 & 9,365 & 14,708 & 12,228 \\
      Movies and TV (Mov.) & 5,289 & 10,138 & 617,122 & 17,732 & 14,804 & 14,669 \\
      Cell Phones and Accessories (Cel.) & 39,262 & 19,868 & 1,211,712 & 63,152 & 97,097 & 100,437 \\
      Sports and Outdoors (Spo.) & 34,571 & 30,506 & 1,468,421 & 87,015 & 113,283 & 102,969 \\
      Books (Boo.) & 38,418 & 34,642 & 2,578,864 & 100,263 & 90,801 & 95,406 \\
      Health and Household (Hea.) & 113,393 & 57,727 & 3,692,658 & 252,223 & 304,222 & 290,457 \\ \bottomrule
      \end{tabular}
\end{table}

\noindent\textbf{Dataset Splits.} To align with our two focal scenarios, we employ the \textit{Amazon} dataset\footnote{\url{https://amazon-reviews-2023.github.io/}}, treating different categories as distinct domains and partitioning the data strictly by timestamp into four temporal periods~\citep{jiCriticalStudy2023}. We utilize six domains with varying scales, sparsity levels, and semantic correlations. The detailed data statistics is presented in Table~\ref{tab:data_split}. Specifically: 
\begin{enumerate}[leftmargin=18pt]
    \item Data before \textit{July 1, 2022} ($t_0$) is used for \textbf{pretraining} to establish the model backbone;
    \item Two incremental training phases, \textit{July-September 2022} ($t_0 \rightarrow t_1$, denoted as $P_1$) and \textit{October-December 2022} ($t_1 \rightarrow t_2$, denoted as $P_2$), are used to train two successive checkpoints;
    \item The final checkpoint at $t_2$ is evaluated on interactions occurring between \textit{January-March 2023}.
\end{enumerate}

\begin{figure*}[!t]
    \centering
    \includegraphics[width=\linewidth]{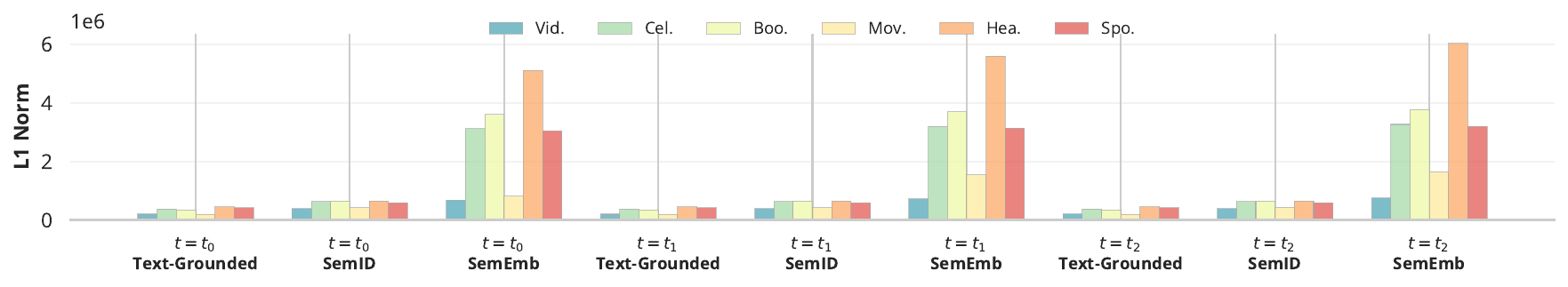}
    \caption{Task vector magnitude of fine-tuned model checkpoints with the same base model \textit{Qwen3-0.6B}.}
    \Description{
        This figure shows the $\mathcal{L}_1$ norms of task vectors across the three GR paradigms for different domains.
    }
    \label{fig:task_vector_magnitude}
\end{figure*}

\noindent\textbf{Models.} We investigate three representative GR approaches in each paradigm defined in Section \ref{ssec:gr} that adopt different paradigms for integrating LLMs with recommendation:
\begin{itemize}[leftmargin=12pt]
\item \textbf{Text-Grounded (BIGRec)}~\citep{baoBiStepGrounding2025} directly leverages textual item information (e.g., titles) to construct sequential recommendation samples for LLM fine-tuning. During inference, the LLM's textual predictions are encoded using an embedding model, followed by retrieval from the item corpus.
\item \textbf{SemID (LC-Rec)}~\citep{zhengAdaptingLarge2024} constructs a series of LLM fine-tuning tasks to align semantic indexes with semantic information, including semantic index-based sequential recommendation and translation tasks between semantic information and indexes. Semantic indexes are generated by training a RQ-VAE~\citep{zeghidourSoundStreamEndtoEnd2021} model and encoded from the item semantic information. The semantic indexes are then treated as expanded vocabulary for the LLM, and be used to represent the item during training and inference.
\item \textbf{SemEmb (HLLM)}~\citep{chenHLLMEnhancing2024} implements a two-stage approach. First, an item LLM is fine-tuned to encode item semantic information based on an LLM backbone. Subsequently, a user LLM takes user history as an item embedding sequence and predicts the next item embedding. Both models are jointly trained during the fine-tuning stage, constituting the complete HLLM framework.
\end{itemize}

\begin{table*}[h]
    \centering
    \renewcommand{\arraystretch}{0.6}
    \scriptsize
    \setlength\tabcolsep{2pt}
    \setul{1pt}{0.4pt}
    \caption{Performance comparison of different GR paradigms on six recommendation domains.}
    \label{tab:baseline_results}
    \begin{tabular}{@{}crrrrrrrrrrrrrrrrrrrrrrrr@{}}
      \toprule
      \textbf{Domain} & \multicolumn{4}{c}{\textbf{Vid.}} & \multicolumn{4}{c}{\textbf{Mov.}} & \multicolumn{4}{c}{\textbf{Cel.}} & \multicolumn{4}{c}{\textbf{Spo.}} & \multicolumn{4}{c}{\textbf{Boo.}} & \multicolumn{4}{c}{\textbf{Hea.}} \\ \midrule
      Metric & \multicolumn{1}{c}{R@10} & \multicolumn{1}{c}{R@20} & \multicolumn{1}{c}{N@10} & \multicolumn{1}{c}{N@20} & \multicolumn{1}{c}{R@10} & \multicolumn{1}{c}{R@20} & \multicolumn{1}{c}{N@10} & \multicolumn{1}{c}{N@20} & \multicolumn{1}{c}{R@10} & \multicolumn{1}{c}{R@20} & \multicolumn{1}{c}{N@10} & \multicolumn{1}{c}{N@20} & \multicolumn{1}{c}{R@10} & \multicolumn{1}{c}{R@20} & \multicolumn{1}{c}{N@10} & \multicolumn{1}{c}{N@20} & \multicolumn{1}{c}{R@10} & \multicolumn{1}{c}{R@20} & \multicolumn{1}{c}{N@10} & \multicolumn{1}{c}{N@20} & \multicolumn{1}{c}{R@10} & \multicolumn{1}{c}{R@20} & \multicolumn{1}{c}{N@10} & \multicolumn{1}{c}{N@20} \\ \midrule
      SASRec & 2.707 & 4.081 & 1.407 & 1.753 & 1.418 & 1.853 & 0.850 & 0.957 & 1.026 & 1.493 & 0.534 & 0.651 & 0.385 & 0.642 & 0.193 & 0.258 & 1.395 & 1.994 & 0.795 & 0.944 & 0.696 & 1.026 & 0.382 & 0.464 \\
      Text-Grounded & 1.723 & 2.707 & 0.897 & 1.143 & 1.560 & 2.180 & 0.760 & 0.919 & 0.660 & 1.120 & 0.320 & 0.435 & 0.240 & 0.400 & 0.124 & 0.164 & 1.540 & 2.220 & 0.784 & 0.955 & 0.480 & 0.800 & 0.262 & 0.342 \\
      SemID & 2.953 & 4.594 & 1.746 & 2.158 & 1.460 & 2.220 & 0.809 & 0.998 & 1.820 & 2.620 & 0.980 & 1.181 & 1.020 & 1.440 & 0.531 & 0.638 & 1.480 & 2.420 & 0.777 & 1.012 & 1.680 & 2.260 & 0.886 & 1.027 \\
      SemEmb & 4.163 & 6.542 & 2.225 & 2.821 & 2.704 & 4.216 & 1.304 & 1.688 & 3.293 & 4.908 & 1.775 & 2.183 & 1.851 & 2.916 & 0.964 & 1.230 & 4.711 & 7.054 & 2.528 & 3.118 & 2.020 & 3.174 & 1.060 & 1.349 \\ \bottomrule
      \end{tabular}
  \end{table*}

\noindent\textbf{Implementation Details.} All three methods are trained using LLMs as backbone models but do not depend on specific model architectures. Here we employ the unified decoder-only \textit{Qwen3-0.6B} model~\citep{yangQwen3Technical2025} as the backbone, consistent with the original papers and common practice in GR regarding model scale~\citep{dengOneRecUnifying2025,zhaiActionsSpeak2024}. All training stages adopts the same learning objectives using AdamW optimizer with fixed learning rates and epoch number (3 epoches in pretraining and 2 epoches in phase 1 and phase 2) for all paradigms, domains and checkpoints. For the embedding model used in BIGRec and LC-Rec, we use \textit{Qwen3-Embedding-8B}~\citep{zhangQwen3Embedding2025}. For LC-Rec, following settings of the original work, we configure 4 layers with 256 tokens per layer as the vocabulary size of semantic IDs.
For HLLM, since the original results indicate that User LLM does not exhibit significant scaling laws~\citep{chenHLLMEnhancing2024}, we adopt the original approach and utilize first and last decoder layers from \textit{Qwen3-0.6B} as the user LLM to ensure comparable model scales across the three methods.

\subsection{Baseline Performance}
The performance of the individually fine-tuned, un-merged models (our baselines) is presented in Table~\ref{tab:baseline_results}. Here we use the last checkpoint ($t_2$) of each domain trained in Section \ref{sec:setup} as the baseline. For comparison, traditional sequential recommendation model SASRec~\citep{kangSelfAttentiveSequential2018} are also included as reference baselines.
We observe several notable patterns in the baseline results. First, \textbf{the text-grounded approach (BIGRec) generally underperforms compared to the other GR methods}. This is likely attributable to its knowledge representation, which is constrained to the textual domain. This makes it difficult to leverage the rich collaborative signals captured by ID-based or embedding-based approaches.
Second, we note a \textbf{consistent performance gap between HLLM and LC-Rec}, with HLLM demonstrating superior performance across domains. We hypothesize this advantage stems from their differing architectural designs: HLLM's jointly optimizes semantic representations alongside collaborative signals, enabling the item embeddings to be directly informed by collaborative patterns. In contrast, LC-Rec adopts a two-stage approach, where semantic IDs are generated without direct access to collaborative information during the codebook learning phase. Although several recent works have proposed improvements to address this limitation~\citep{liSemanticConvergence2025}, we retain the original LC-Rec implementation as a representative, since our primary objective is to explore the behavior of different types of GR methods under model merging rather than their absolute performance. This allows us to systematically investigate how different recommendation generation mechanisms respond to merging operations across our identified scenarios.

\section{Creating Unified GR via Cross-domain Merging: Challenges, Causes and Solutions}
\label{sec:cross_domain}
In this section, following the emerging trend of developing unified GR models, we elaborate on the challenges faced when creating a cross-domain GR through MM. We systematically discuss key difficulties and potential sources that affect the performance of cross-domain merging, and provide potential practical solutions to address these issues.

\subsection{Initial Experiments and Analysis}
To explore the potential of MM on cross-domain GR, we first adopt the most fundamental setting: merging between two domains and evaluating performance of the merged model on each source domain separately.

\paragraph{Experimental design.} Following Definition \ref{def:cross_domain_merging}, we leverage the last checkpoint ($t_2$) of each domain trained in Section \ref{sec:setup}, and merge the two model weights $\Phi_{\Theta^{(\mathcal{D}_i, t_2)}}$ and $\Phi_{\Theta^{(\mathcal{D}_j, t_2)}}$ from domains $\mathcal{D}_i$ and $\mathcal{D}_j$, generating a unified model $\Theta^{*}$. Following the task arithmetic framework~\citep{ilharcoEditingModels2022}, we compute task vectors for all fine-tuned models as $\tau(\Theta^{(0)}, \Theta^{(k)}) = \Theta^{(k)} - \Theta^{(0)}$, where $\Theta^{(0)}$ denotes the original \textit{Qwen3-0.6B} base model. Two widely-used merging methods are employed: \textbf{Weighted-based} and \textbf{Subspace-based} merging. Specifically, we set a equal weight $\lambda = 0.5$ for the weighted-based merging~\citep{ilharcoEditingModels2022}, and adopt both DARE ($p=0.1$)~\citep{yangModelMerging2024} and TIES ($X=20$)~\citep{yadavTIESMergingResolving2023} simutaneously for the subspace-based merging.
\footnote{For the SemID method, since item semantic IDs are generated independently across domains, establishing one-to-one parameter correspondence for merging is infeasible. Therefore, we retain the token embeddings from both domains and incorporate them into the merged model's expanded vocabulary.}

\begin{figure}[!t]
    \centering
    \includegraphics[width=0.85\linewidth]{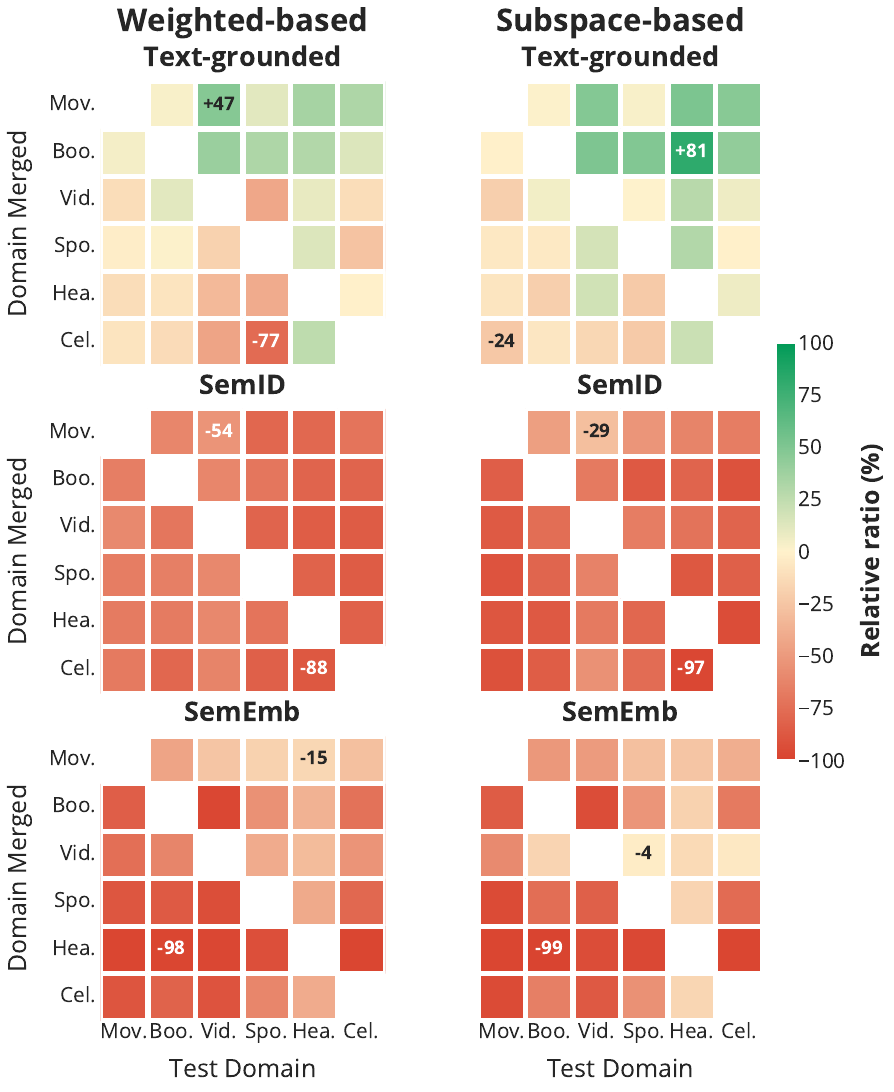}
    \caption{Performance heatmap of cross-domain merging. }
    \Description{
        The performance heatmap of cross-domain merging. The x-axis represents the source domain, and the y-axis represents the target domain. The color represents the performance improvement of the merged model compared to the source model.
    }
    \label{fig:PerformanceHeatmap}
\end{figure}

\paragraph{Performance variation analysis.} To explicitly visualize the relative performance variations of different domains and GR paradigms, we set the performance of single domain model as the baseline, and normalize the ratio of performance change of the merged model to the baseline. The percentage improvement or degradation of the merged model compared to the baseline on each domain is visualized as heatmaps in Figure~\ref{fig:PerformanceHeatmap}, revealing several critical findings:

\textbf{Finding 1: Text-Grounded GR shows semantic correlated positive transfer.} In the initial experiments, Text-grounded is the only GR paradigm that achieves performance improvements through MM, and the magnitude of improvement appears correlated with the semantic characteristics of domains. Domains with \textbf{diverse semantic knowledge}, such as Books and Movies, tend to provide greater benefits when merged with other domains. Domain pairs with totally different semantic characteristics, such as Cellphones and Sports, tend to show performance degradation. This suggests that text-grounded representations can effectively leverage cross-domain semantic commonalities encoded in the LLM's pretrained knowledge.

\textbf{Finding 2: SemID and SemEmb paradigms exhibit substantial domain conflicts.} Both SemID and SemEmb models demonstrate widespread and significant performance declines, with degradation severity varying by cross-domain merging.
We hypothesize that this degradation stems from fundamental differences in how these paradigms modify the pretrained LLM. Unlike text-grounded approaches where both input and output remain fully textual and naturally align with the LLM's original distribution, SemID introduces \textbf{out-of-distribution tokens} into the vocabulary, while SemEmb learns a fundamentally \textbf{different task} that encodes collaborative signals into dense item representations. These architectural choices amplify parameter shifts from the pretrained base model, making the resulting task vectors highly task-dependent. Consequently, the out-of-distribution nature of semantic IDs and the entangled user-item collaborative signals in embedding spaces render these representations particularly susceptible to destructive interference during parameter merging, leading to the observed performance degradation.

\textbf{Finding 3: SemEmb exhibits unbalanced performance variation correlated with dataset size.} 
Compared to the other two paradigms, SemEmb demonstrates the most pronounced asymmetry in cross-domain merging outcomes. As shown in Figure~\ref{fig:PerformanceHeatmap} and Table~\ref{tab:data_split}, domains with larger training datasets experience minimal performance degradation after merging, while domains with smaller training datasets suffer substantial performance drops. This imbalance suggests that the merged model's parameters and task vectors are disproportionately influenced by the domain with more training data, effectively dominating the shared parameter space and marginalizing the contributions from smaller domains.

\subsection{Observing Domain Conflict via Task Vectors}
To further investigate the root causes of performance degradation in SemID and SemEmb paradigms, we analyze the task vectors across different domains and measure their magnitudes. 
Figure~\ref{fig:task_vector_magnitude} illustrates the $\mathcal{L}_1$ norms of task vectors across the three GR paradigms for different domains. We observe a clear hierarchy: SemEmb exhibits the largest magnitude changes, followed by SemID, with Text-grounded showing the smallest deviations. Meanwhile, compared to SemID and Text-grounded, a significant magnitude variation across domains is observed for SemEmb. 

\begin{figure*}[!t]
    \centering
    \includegraphics[width=0.85\linewidth]{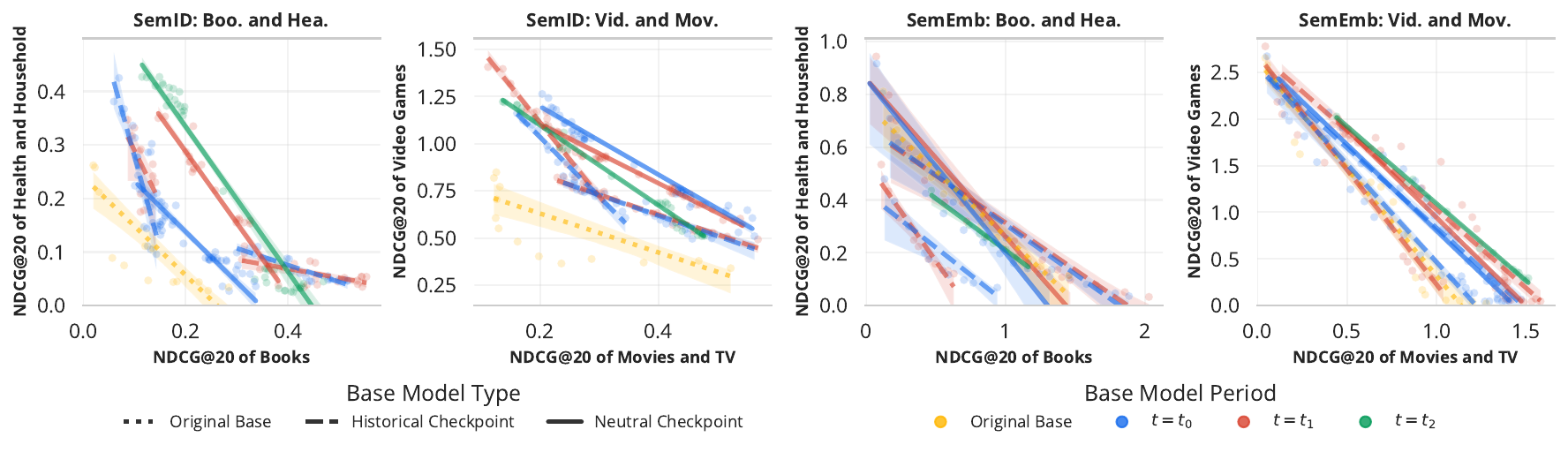}
    \caption{Joint performance comparsion of cross-domain merging with different base models. }
    \Description{
        This figure shows the performance comparison of different base models. The x-axis represents the source domain, and the y-axis represents the target domain. The color represents the performance improvement of the merged model compared to the source model.
    }
    \label{fig:base_comparison}
\end{figure*}

From the above observations, we hypothesize that the performance degradation in SemID and SemEmb paradigms stems from several interrelated factors:

First, compared to Text-grounded approaches, SemID introduces \textbf{out-of-distribution tokens} into the vocabulary, creating a fundamental mismatch with the pretrained LLM's original token distribution. This architectural modification forces the model to learn representations for tokens that lie outside the natural language space the LLM was originally trained on.

Second, compared to SemID, SemEmb exhibits greater magnitude variation across domains, which is likely caused by \textbf{learning a fundamentally different task compared with the original auto-regressive generation objective}. While SemID maintains the generation framework but with additional tokens, SemEmb transforms the task into the contrastive learning of collaborative embeddings, resulting in more substantial parameter shifts growing with the number of training data.

Third, both SemID and SemEmb show relatively stable magnitude across checkpoints at $t_1$ and $t_2$, indicating that \textbf{the task-aware parameter shifts to accommodate distribution adaptation and objective changes occur predominantly during the initial pretraining phase}. This observation reveals that \textbf{task-aware parameter shifts can potentially be disentangled from the task vectors}, enabling the computed task vectors to more specifically reflect domain-incremental knowledge rather than conflating task adaptation.

\subsection{Disentangling Task-Aware Parameter Shift}
Based on the hypothesis that task vectors contain excessive task-specific information, we propose an approach to mitigate conflicts of cross-domain merging in GR by replacing the base model used for task vector computation. Our goal is to identify a base model that already encapsulates substantial task-aware but domain-independent general recommendation knowledge, such that the computed task vectors $\tau^{\prime}$ represent purely domain-incremental knowledge.

\paragraph{Leveraging historical checkpoints as base model.} In continual learning scenarios for RSs, we possess checkpoints trained on the same domain $\mathcal{D}$ at different time points $t_1 < t_2$. Leveraging the property that $\Theta^{(\mathcal{D}, t_1)}$ has already been fine-tuned and adapted to the recommendation task, we attempt to use it as the new base model $\Theta^{(0)\prime} = \Theta^{(\mathcal{D}, t_1)}$, replacing the original pretrained model $\Theta^{(0)}$. We recompute task vectors based on the new base $\Theta^{(0)\prime}$ as $\tau^{\prime} = \Theta^{(\mathcal{D}, t_2)} - \Theta^{(0)\prime}$, and perform cross-domain merging using these new task vectors. In cross-domain merging scenario, we use the checkpoint from one of the domains to merge as the base model.

\paragraph{Experimental details.}
To validate the effectiveness of the base model replacement strategy, we conduct experiments with the replaced base model. Here we adopt the subspace-based merging method in Figure~\ref{fig:PerformanceHeatmap} because it shows less performance degradation in cross-domain merging.
However, the phenomenon in Figure \ref{fig:task_vector_magnitude} suggests a potential challenge: the task vectors from different domains may become dominant due to differences in magnitude. 
To mitigate its effect, we introduce an adjustable factor $\alpha$ to rescale the task vector, and visualize the joint performance of cross-domain merging with different $\alpha$ values.
\begin{equation}
    \small
    \Theta^{*} =  \alpha \textsc{Trim}(\boldsymbol{\tau}(\Theta^{(0)'}, \Theta^{(\mathcal{D}_i, t_2)})) + (1-\alpha) \textsc{Trim}(\boldsymbol{\tau}(\Theta^{(0)'}, \Theta^{(\mathcal{D}_j, t_2)})).
\end{equation}

We visualize the joint performance of cross-domain merging with different $\alpha$ values in Figure~\ref{fig:base_comparison}.
The results show that replaced base model (marked as "Historical Checkpoint") improves merging effectiveness, especially for SemID, validating the efficacy of disentangling general task knowledge. However, using a historical checkpoint from one domain as the base model inherently biases the merged result toward that domain, as the base model already encodes substantial domain-specific knowledge, making it difficult to achieve balanced cross-domain performance.

\paragraph{Creating neutral base model.} To address this limitation, we propose to create a base model that is more "neutral" and "task-aware" for all domains participating in the merge. To obtain a base model that is more "neutral" and "task-aware" for all domains participating in the merge, we attempt to combine prior knowledge from both domains.
We adopt the simple average merging to combine the checkpoints at the same training phase from both domains to construct a neutral base:
\begin{equation}
    \small
    \Theta^{(\mathcal{D}_i \cup \mathcal{D}_j, t)} = \textsc{Average}(\Theta^{(\mathcal{D}_i, t)}, \Theta^{(\mathcal{D}_j, t)}).
\end{equation}
Task vectors for models at $t_2$ are then recomputed based on $\Theta^{(\mathcal{D}_i \cup \mathcal{D}_j, t)}$ before merging. Figure~\ref{fig:base_comparison} presents the merging performance. Results demonstrate further performance improvements with the neutral base strategy, suggesting that this approach more effectively separates domain-specific knowledge, thereby reducing redundant task-aware information in task vectors. It is worth noting that similar strategies have been discussed in the literature of MM~\citep{choiRevisitingWeight2025}, which corroborates our conclusions.

\paragraph{Limitations and potential for further improvement.} Meanwhile, we also notice that the performance is still far from that of each domain-specific checkpoint. For domains with large parameter shifts (such as Books and Health domains in SemEmb), the simple averaged neutral base model brought limited performance gain. This suggests that there is still potential to develop more sophisticated strategies to better disentangle the task-aware parameter shift from the task vectors, aiming to create a base model to derive task vectors more specific to domain knowledge.

\section{Balancing Evolving User Preferences via Temporal Merging}
\label{sec:temporal}
In this section, we are seeking to exploit the potential of MM in \textbf{incremental training} scenraio of practical RSs.
In traditional applications of MM, merging checkpoints from the training process to facilitate fine-tuning is a widely-adopted practice~\citep{liuLinearCombination2024}. While this shares similarities with continual training in RSs, a critical distinction exists: \textbf{continual training data in RSs inherently encodes the temporal variance of distribution in the training data}. In incrementally trained RSs, newly incorporated training data originates from fresh user interactions, typically encoding recent preference shifts among currently active users~\citep{guoEnhancingNewitem2025}. This suggests that parameter changes between old and new checkpoints may \textbf{implicitly capture trends in evolving user preferences}, hinting at the possibility of leveraging these trends to fine-tune RSs and balance historical and emerging preferences to achieve better overall performance.

Consider the temporal dynamics of user behavior: early checkpoints trained on historical data may excel at capturing stable, long-term user interests, while recent checkpoints trained on fresh interactions are more attuned to emerging trends and shifting preferences. However, the most recent checkpoint alone may \textbf{overfit to transient patterns} or suffer from \textbf{recency bias}~\citep{chenCorrectingRecency2019}, neglecting valuable stable preferences encoded in earlier training phases. This motivates us to explore whether strategic merging of \textbf{temporal checkpoints} can synthesize the complementary strengths of different training stages, preserving robust historical knowledge while incorporating emerging behavioral patterns.

In this section, we investigate how to perform such optimization through MM of temporal checkpoints. Specifically, we examine whether merging models \textbf{from different time periods} can yield a unified model that outperforms either individual checkpoint by effectively balancing temporal dynamics in user preferences. 

\subsection{Constructing Task Vectors to Reflect Temporal Preference Shift}
Following the task vector formulation, we hypothesize that the parameter delta between checkpoints trained on data from different time periods encodes the \textbf{temporal preference shift} occurring during that interval. Formally, we define:

\begin{definition}[Temporal Preference Shift Vector]
Given two checkpoints $\Theta^{(\mathcal{D}, t_1)}$ and $\Theta^{(\mathcal{D}, t_2)}$ trained sequentially on domain $\mathcal{D}$ at times $t_1 < t_2$, the temporal preference shift vector is defined as:
\begin{equation}
    \tau_{\text{temp}}^{(\mathcal{D}, t_1, t_2)} = \Theta^{(\mathcal{D}, t_2)} - \Theta^{(\mathcal{D}, t_1)}.
\end{equation}
This vector captures the parameter changes induced by training on incremental data from the time interval $[t_1, t_2]$, encoding emerging user preference patterns and behavioral shifts.
\end{definition}

In this work, we conduct experiments using the difference between our last two training phases: $t_1$ and $t_2$. We treat the checkpoint $\Theta^{(\mathcal{D}, t_1)}$ as the base model and modulate the Temporal Preference Shift Vector with a scaling coefficient $\lambda_{temp}$:
\begin{equation}
    \Theta_{\lambda_{temp}}^{(\mathcal{D}, t_2^*)} = \Theta^{(\mathcal{D}, t_1)} + \lambda_{temp} \cdot \tau_{\text{temp}}^{(\mathcal{D}, t_1, t_2)}.
\end{equation}
When $\lambda_{temp} = 0$, we recover the checkpoint at $t_1$; when $\lambda_{temp} = 1$, we obtain the checkpoint at $t_2$. By varying $\lambda_{temp}$, we can interpolate between historical and recent model states, or even extrapolate beyond the checkpoint at $t_2$ ($\lambda_{temp} > 1$) to amplify emerging preference trends. This formulation allows us to systematically investigate how different weightings of temporal dynamics affect recommendation performance on future user interactions. In this section, we choose SemEmb models for the experiments due to the its superior performance in the baseline results.

\subsection{Balancing Evolving Preferences via Weighted-based Merging}
\begin{figure}[!t]
    \centering
    \includegraphics[width=0.95\linewidth]{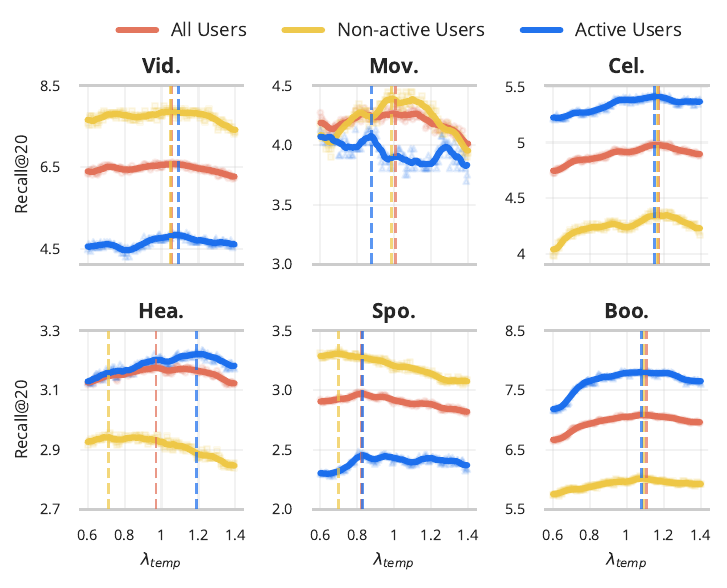}
    \caption{Temporal preference shift of fine-tuned models. The vertical dashed line indicates the $\lambda_{temp}$ value that achieves the best performance for each domain and user group.}
    \Description{
        This figure shows the performance of temporal preference shift of fine-tuned models. The x-axis represents the value of $\lambda_{temp}$, and the y-axis represents the performance.
    }
    \label{fig:lambda_performance}
\end{figure}

\begin{table*}[!t]
    \centering
    \renewcommand{\arraystretch}{0.7}
    \scriptsize
    \setlength\tabcolsep{2pt}
    \setul{1pt}{0.4pt}
    \caption{Performance comparison with different $\lambda_{temp}$ values.}
    \label{tab:lambda_performance_comparison}
    \begin{tabular}{@{}lcccccccccccccccccccccccc@{}}
        \toprule
        \multicolumn{1}{c}{\textbf{Domain}} & \multicolumn{4}{c}{\textbf{Vid.}} & \multicolumn{4}{c}{\textbf{Mov.}} & \multicolumn{4}{c}{\textbf{Cel.}} & \multicolumn{4}{c}{\textbf{Spo.}} & \multicolumn{4}{c}{\textbf{Boo.}} & \multicolumn{4}{c}{\textbf{Hea.}} \\ \midrule
        \multicolumn{1}{c}{\textbf{Metric}} & R@10 & R@20 & N@10 & N@20 & R@10 & R@20 & N@10 & N@20 & R@10 & R@20 & N@10 & N@20 & R@10 & R@20 & N@10 & N@20 & R@10 & R@20 & N@10 & N@20 & R@10 & R@20 & N@10 & N@20 \\ \midrule
        $\lambda_{temp}=1$ & 4.163 & 6.542 & \textbf{2.225} & \textbf{2.821} & 2.704 & 4.216 & 1.304 & 1.688 & 3.293 & 4.908 & 1.775 & 2.183 & 1.851 & 2.916 & 0.964 & 1.230 & 4.711 & 7.054 & \textbf{2.528} & \textbf{3.118} & \textbf{2.020} & \textbf{3.174} & \textbf{1.060} & \textbf{1.349} \\
        $\lambda_{temp}=\hat{\lambda}_{temp}^{*}$ & \textbf{4.225} & \textbf{6.604} & 2.204 & 2.800 & \textbf{2.798} & \textbf{4.235} & \textbf{1.353} & \textbf{1.715} & \textbf{3.321} & \textbf{4.931} & \textbf{1.781} & \textbf{2.187} & \textbf{1.866} & \textbf{2.956} & \textbf{0.968} & \textbf{1.240} & \textbf{4.724} & \textbf{7.062} & 2.523 & 3.109 & 1.966 & 3.155 & 1.040 & 1.340 \\ \bottomrule
        \end{tabular}
\end{table*}

We first visualize the performance variations across different values of $\lambda_{temp}$ in Figure~\ref{fig:lambda_performance} (denoted as "All Users"). First, a striking observation emerges: $\lambda_{temp} = 1 $ \textbf{does not consistently yield optimal performance across domains}. This suggests that simply training on the most recent interaction data may fail to achieve the optimal balance between historical and emerging user preferences in RSs.  Meanwhile, it is also observed that the optimal $\lambda_{temp}$ value varies across domains, suggesting that balancing the historical and emerging user preferences is a domain dependent problem.

To further investigate the domain-dependent nature of temporal preference shift, we partition the test set into two groups based on the user's interaction history during the time interval $[t_1, t_2]$:
\begin{itemize}[leftmargin=12pt]
    \item \textbf{Active Users}: Users who performed at least one interaction during the time interval $[t_1, t_2]$.
    \item \textbf{Non-Active Users}: Users with no interactions during the time interval $[t_1, t_2]$.
\end{itemize}
Then, we measure the performance of these two user groups as $\lambda_{temp}$ varies. As shown in Figure~\ref{fig:lambda_performance}, active users generally show smaller variance in the optimal value of $\lambda_{temp}$ that achieves the best performance. This is intuitive since the incremental training data directly reflects their recent behavioral patterns.
However, non-active users exhibit dramatically different trends. These users' preferences were encoded in the earlier checkpoint $\Theta^{(\mathcal{D}, t_1)}$. Since the training data lacks their recent interactions, the parameter shifts induced by this new data may \textbf{positively or negatively affect their recommendations in unpredictable ways}, causing the optimal $\lambda_{temp}$ value to vary substantially across domains.
Notably, if non-active users experience preference shifts similar to active users, even without explicitly contributing interactions, the parameter changes from incremental training data could still benefit them. This observation motivates us to investigate domain characteristics that might correlate with collective preference shifting patterns.

\paragraph{Investigating domain characteristics.} Inspired by the above observations, we delve deeper into domain data characteristics to identify factors potentially associated with preference shifting. RSs operate in dynamic environments where both users and items continuously evolve. \textbf{The influx of new items may be a critical driver of preference shifts}. In Figure~\ref{fig:lambda_performance}, the \textit{Cell Phones} domain exhibits remarkably similar preference shift patterns between active and non-active users. Given the rapid iteration cycles of smartphones and accessories, this aligns with our intuition that both active and non-active users tend to gravitate toward newer products, inducing similar preference shifts across the user base.

\paragraph{Item recency analysis.} To validate this hypothesis, we compute the time gap between each interaction timestamp and the item's first appearance time in our dataset. Since our dataset lacks explicit release dates, we use the timestamp of an item's first interaction as a proxy. Table~\ref{tab:item_recency_correlation} shows the statistics of the time gaps across domains in training sets, together with the optimal $\lambda_{temp}*$ values (denoted as $\lambda_{temp}^{*}$) for non-active users in each domain.

\begin{table}[!t]
    \centering
    \renewcommand{\arraystretch}{0.7}
    \footnotesize
    \setlength\tabcolsep{3pt}
    \setul{1pt}{0.4pt}
    \caption{Statistics and analysis of item recency bias.}
    \label{tab:item_recency_correlation}
    \begin{tabular}{@{}lrrrrrr@{}}
        \toprule
        \textbf{Domain} & \multicolumn{1}{c}{\textbf{Hea.}} & \multicolumn{1}{c}{\textbf{Spo.}} & \multicolumn{1}{c}{\textbf{Mov.}} & \multicolumn{1}{c}{\textbf{Vid.}} & \multicolumn{1}{c}{\textbf{Boo.}} & \multicolumn{1}{c}{\textbf{Cel.}} \\ \midrule
        \textbf{\# Users} & 113,393 & 34,571 & 5,289 & 4,876 & 38,418 & 39,262 \\
        \textbf{\# Active Users} & 44,725 & 15,525 & 1,839 & 2,202 & 17,589 & 15,647 \\
        \textbf{\# Non-active Users} & 68,668 & 19,046 & 3,450 & 2,674 & 20,829 & 23,615 \\
        \textbf{Ratio of Active Users} in $P_2$ & 0.394 & 0.449 & 0.348 & 0.452 & 0.458 & 0.399 \\ \midrule
        \textbf{$\lambda_{temp}^{*}$ (Non-active Users)} & 0.715 & 0.700 & 0.985 & 1.050 & 1.090 & 1.165 \\
        \textbf{Ave. Interaction Time Gap (d)} & 662.9 & 563.9 & 520.8 & 418.2 & 336.6 & 362.0 \\ \bottomrule
        \end{tabular}
\end{table}

Remarkably, a strong correlation emerges: domains where users interact tend to interact with newer items, corresponding to smaller optimal $\lambda_{temp}$ values for non-active users, suggesting greater benefit from recent model updates. Conversely, domains with longer item lifespans require larger $\lambda_{temp}$ values, favoring historical knowledge. This finding demonstrates that \textbf{domain-specific temporal characteristics can guide the selection of merging weights for temporal preference shift vectors}, enabling better balance between historical and emerging preferences. Critically, these domain statistics can be computed from training data alone, allowing us to transfer insights across domains and inform weighted merging strategies without requiring labeled test data.

\paragraph{Predicting optimal $\lambda_{temp}^{*}$ via domain characteristics.}
To validate the correlation between the domain characteristics and the optimal $\lambda_{temp}^{*}$, we conduct a simple experiment, using the average interaction time gap and the optimal $\lambda_{temp}^{*}$ to learn a \textbf{linear regression model}. 
A leave-one-out strategy is adopted to train the regression model with data from five domains and evaluate the performance on the left-out domain. 
As the prediction results is expected to be the optimal $\lambda_{temp}$ value for non-active users, we suppose a well-trained model should align with active users' preference at the $\lambda_{temp}=1$ value that achieves the best performance. Then, we simply use the weighted averge by the ratio of active users (denoted as $p_{active}$) to compute the final predicted optimal $\lambda_{temp}$ value as $\hat{\lambda}_{temp}^{*}$:
\begin{equation}
    \hat{\lambda}_{temp}^{*} = (1 - p_{active})  \cdot \lambda_{temp}^{*} + p_{active} \cdot 1.
\end{equation}
The overall performance compared with the original checkpoint ($\lambda_{temp}=1$) is shown in Table~\ref{tab:lambda_performance_comparison}. It is observed that the performance of the predicted $\lambda_{temp}$ value is generally better than the original checkpoint, suggesting that the proposed approach can effectively balance the historical and emerging user preferences.

\paragraph{Limitations and potential for further improvement.} The above preliminary analysis demonstrates the potential of MM in balancing historical and emerging user preferences in the context of incremental training for RSs. By leveraging cross-domain statistics, we can effectively guide the merging process to achieve better performance than naive approaches. However, several limitations remain. First, the linear regression model, while simple and interpretable, \textbf{may not capture complex nonlinear relationships between domain characteristics and optimal merging weights}. Second, our analysis focuses primarily on item recency as a domain characteristic, but other factors such as \textbf{user engagement patterns}, item diversity, and seasonal trends may also play important roles.


\section{Conclusion}
In this work, we introduced \frameworkname, a unified framework for studying model merging in generative recommendation from a contextual perspective.
By organizing context-specialized GR checkpoints into a structured model space, the contextual grid enables systematic analysis of merging behaviors under realistic deployment conditions with heterogeneous domains and evolving nature.
Through extensive experiments, we revealed key challenges and insights regarding contextual conflicts and imbalance in model merging, and demonstrated effective strategies to mitigate them.
Our findings highlight the critical role of contextual characteristics in shaping merging effectiveness and provide practical guidance for designing robust merging strategies for GR.
We believe \frameworkname will serve as a valuable foundation for advancing both the theoretical understanding and practical application of model merging in real-world generative recommender systems.


\bibliographystyle{ACM-Reference-Format}
\bibliography{MMGenRec}

\appendix



\end{document}

%% file: Main/authors.tex
\author{Tianjun Wei}
\orcid{0000-0001-7311-7101}
\affiliation{%
  \institution{Nanyang Technological University}
  \country{Singapore}
  \city{Singapore}
}
\email{tjwei2-c@my.cityu.edu.hk}

\author{Enneng Yang}
\orcid{0000-0001-5419-5286}
\affiliation{%
  \institution{Nanyang Technological University}
  \country{Singapore}
  \city{Singapore}
}
\email{N2308949L@e.ntu.edu.sg}

\author{Yingpeng Du}
\authornote{Corresponding authors.}
\orcid{0000-0001-9881-7171}
\affiliation{%
  \institution{Nanyang Technological University}
  \country{Singapore}
  \city{Singapore}
}
\email{dyp1993@pku.edu.cn}

\author{Huizhong Guo}
\orcid{0009-0004-0011-8612}
\affiliation{%
  \institution{Zhejiang University}
  \country{China}
  \city{Hangzhou}
}
\email{huiz_g@zju.edu.cn}

\author{Jie Zhang}
\orcid{0000-0001-8996-7581}
\affiliation{%
  \institution{Nanyang Technological University}
  \country{Singapore}
}
\email{zhangj@ntu.edu.sg}

\author{Zhu Sun*}
\orcid{0000-0002-3350-7022}
\affiliation{%
  \institution{Singapore University of Technology and Design}
  \country{Singapore}
}
\email{zhu_sun@sutd.edu.sg}

%% file: MMGenRec.bib
@article{adomaviciusNextGeneration2005,
  title = {Toward the next Generation of Recommender Systems: A Survey of the State-of-the-Art and Possible Extensions},
  shorttitle = {Toward the next Generation of Recommender Systems},
  author = {Adomavicius, G. and Tuzhilin, A.},
  year = 2005,
  month = jun,
  journal = {IEEE Transactions on Knowledge and Data Engineering},
  volume = {17},
  number = {6},
  pages = {734--749},
  issn = {1558-2191},
  doi = {10.1109/TKDE.2005.99},
  urldate = {2025-03-01}
}

@inproceedings{aielloJointlyTraining2023,
  title = {Jointly {{Training Large Autoregressive Multimodal Models}}},
  booktitle = {The {{Twelfth International Conference}} on {{Learning Representations}}},
  author = {Aiello, Emanuele and Yu, Lili and Nie, Yixin and Aghajanyan, Armen and Oguz, Barlas},
  year = 2023,
  month = oct,
  urldate = {2025-11-14},
  langid = {english}
}

@article{akibaEvolutionaryOptimization2025,
  title = {Evolutionary Optimization of Model Merging Recipes},
  author = {Akiba, Takuya and Shing, Makoto and Tang, Yujin and Sun, Qi and Ha, David},
  year = 2025,
  month = feb,
  journal = {Nature Machine Intelligence},
  volume = {7},
  number = {2},
  pages = {195--204},
  publisher = {Nature Publishing Group},
  issn = {2522-5839},
  doi = {10.1038/s42256-024-00975-8},
  urldate = {2025-11-14},
  copyright = {2025 The Author(s)},
  langid = {english}
}

@article{baoBiStepGrounding2025,
  title = {A {{Bi-Step Grounding Paradigm}} for {{Large Language Models}} in {{Recommendation Systems}}},
  author = {Bao, Keqin and Zhang, Jizhi and Wang, Wenjie and Zhang, Yang and Yang, Zhengyi and Luo, Yanchen and Chen, Chong and Feng, Fuli and Tian, Qi},
  year = 2025,
  month = apr,
  journal = {ACM Trans. Recomm. Syst.},
  volume = {3},
  number = {4},
  pages = {53:1--53:27},
  doi = {10.1145/3716393},
  urldate = {2025-11-08}
}

@inproceedings{baoDecodingMatters2024,
  title = {Decoding {{Matters}}: {{Addressing Amplification Bias}} and {{Homogeneity Issue}} in {{Recommendations}} for {{Large Language Models}}},
  shorttitle = {Decoding {{Matters}}},
  booktitle = {Proceedings of the 2024 {{Conference}} on {{Empirical Methods}} in {{Natural Language Processing}}},
  author = {Bao, Keqin and Zhang, Jizhi and Zhang, Yang and Huo, Xinyue and Chen, Chong and Feng, Fuli},
  editor = {{Al-Onaizan}, Yaser and Bansal, Mohit and Chen, Yun-Nung},
  year = 2024,
  month = nov,
  pages = {10540--10552},
  publisher = {Association for Computational Linguistics},
  address = {Miami, Florida, USA},
  doi = {10.18653/v1/2024.emnlp-main.589},
  urldate = {2025-11-08}
}

@inproceedings{chenCorrectingRecency2019,
  title = {Correcting for {{Recency Bias}} in {{Job Recommendation}}},
  booktitle = {Proceedings of the 28th {{ACM International Conference}} on {{Information}} and {{Knowledge Management}}},
  author = {Chen, Ruey-Cheng and Ai, Qingyao and Jayasinghe, Gaya and Croft, W. Bruce},
  year = 2019,
  month = nov,
  series = {{{CIKM}} '19},
  pages = {2185--2188},
  publisher = {Association for Computing Machinery},
  address = {New York, NY, USA},
  doi = {10.1145/3357384.3358131},
  urldate = {2025-12-25},
  isbn = {978-1-4503-6976-3}
}

@misc{chenHLLMEnhancing2024,
  title = {{{HLLM}}: {{Enhancing Sequential Recommendations}} via {{Hierarchical Large Language Models}} for {{Item}} and {{User Modeling}}},
  shorttitle = {{{HLLM}}},
  author = {Chen, Junyi and Chi, Lu and Peng, Bingyue and Yuan, Zehuan},
  year = 2024,
  month = sep,
  number = {arXiv:2409.12740},
  eprint = {2409.12740},
  primaryclass = {cs},
  publisher = {arXiv},
  doi = {10.48550/arXiv.2409.12740},
  urldate = {2025-11-08},
  archiveprefix = {arXiv}
}

@misc{choiRevisitingWeight2025,
  title = {Revisiting {{Weight Averaging}} for {{Model Merging}}},
  author = {Choi, Jiho and Kim, Donggyun and Lee, Chanhyuk and Hong, Seunghoon},
  year = 2025,
  month = apr,
  number = {arXiv:2412.12153},
  eprint = {2412.12153},
  primaryclass = {cs},
  publisher = {arXiv},
  doi = {10.48550/arXiv.2412.12153},
  urldate = {2025-12-24},
  archiveprefix = {arXiv}
}

@inproceedings{dengLogitSpace2025,
  title = {Logit {{Space Constrained Fine-Tuning}} for {{Mitigating Hallucinations}} in {{LLM-Based Recommender Systems}}},
  booktitle = {Proceedings of the 2025 {{Conference}} on {{Empirical Methods}} in {{Natural Language Processing}}},
  author = {Deng, Jianfeng and Chen, Qingfeng and Cheng, Debo and Li, Jiuyong and Liu, Lin},
  editor = {Christodoulopoulos, Christos and Chakraborty, Tanmoy and Rose, Carolyn and Peng, Violet},
  year = 2025,
  month = nov,
  pages = {29299--29312},
  publisher = {Association for Computational Linguistics},
  address = {Suzhou, China},
  urldate = {2025-11-08},
  isbn = {979-8-89176-332-6}
}

@misc{dengOneRecUnifying2025,
  title = {{{OneRec}}: {{Unifying Retrieve}} and {{Rank}} with {{Generative Recommender}} and {{Iterative Preference Alignment}}},
  shorttitle = {{{OneRec}}},
  author = {Deng, Jiaxin and Wang, Shiyao and Cai, Kuo and Ren, Lejian and Hu, Qigen and Ding, Weifeng and Luo, Qiang and Zhou, Guorui},
  year = 2025,
  month = feb,
  number = {arXiv:2502.18965},
  eprint = {2502.18965},
  primaryclass = {cs},
  publisher = {arXiv},
  doi = {10.48550/arXiv.2502.18965},
  urldate = {2025-04-19},
  archiveprefix = {arXiv}
}

@inproceedings{dholeGenerativeProduct2025,
  title = {Generative {{Product Recommendations}} for {{Implicit Superlative Queries}}},
  booktitle = {Proceedings of the 2025 {{Conference}} of the {{Nations}} of the {{Americas Chapter}} of the {{Association}} for {{Computational Linguistics}}: {{Human Language Technologies}} ({{Volume}} 4: {{Student Research Workshop}})},
  author = {Dhole, Kaustubh and Vedula, Nikhita and Kuzi, Saar and Castellucci, Giuseppe and Agichtein, Eugene and Malmasi, Shervin},
  editor = {Ebrahimi, Abteen and Haider, Samar and Liu, Emmy and Haider, Sammar and Leonor Pacheco, Maria and Wein, Shira},
  year = 2025,
  month = apr,
  pages = {77--91},
  publisher = {Association for Computational Linguistics},
  address = {Albuquerque, USA},
  doi = {10.18653/v1/2025.naacl-srw.8},
  urldate = {2025-11-08},
  isbn = {979-8-89176-192-6}
}

@misc{duReinforcementSpeculative2025a,
  title = {Reinforcement {{Speculative Decoding}} for {{Fast Ranking}}},
  author = {Du, Yingpeng and Wei, Tianjun and Sun, Zhu and Zhang, Jie},
  year = 2025,
  month = may,
  number = {arXiv:2505.20316},
  eprint = {2505.20316},
  primaryclass = {cs},
  publisher = {arXiv},
  doi = {10.48550/arXiv.2505.20316},
  urldate = {2026-01-14},
  archiveprefix = {arXiv}
}

@inproceedings{guoEnhancingNewitem2025,
  title = {Enhancing {{New-item Fairness}} in {{Dynamic Recommender Systems}}},
  booktitle = {Proceedings of the 48th {{International ACM SIGIR Conference}} on {{Research}} and {{Development}} in {{Information Retrieval}}},
  author = {Guo, Huizhong and Sun, Zhu and Wang, Dongxia and Wei, Tianjun and Li, Jinfeng and Zhang, Jie},
  year = 2025,
  month = jul,
  series = {{{SIGIR}} '25},
  pages = {1707--1716},
  publisher = {Association for Computing Machinery},
  address = {New York, NY, USA},
  doi = {10.1145/3726302.3729969},
  urldate = {2025-11-14},
  isbn = {979-8-4007-1592-1}
}

@inproceedings{heLLM2RecLarge2025,
  title = {{{LLM2Rec}}: {{Large Language Models Are Powerful Embedding Models}} for {{Sequential Recommendation}}},
  shorttitle = {{{LLM2Rec}}},
  booktitle = {Proceedings of the 31st {{ACM SIGKDD Conference}} on {{Knowledge Discovery}} and {{Data Mining}}},
  author = {He, Yingzhi and Liu, Xiaohao and Zhang, An and Ma, Yunshan and Chua, Tat-Seng},
  year = 2025,
  month = aug,
  series = {{{KDD}} '25},
  pages = {896--907},
  publisher = {Association for Computing Machinery},
  address = {New York, NY, USA},
  doi = {10.1145/3711896.3737029},
  urldate = {2025-12-23},
  isbn = {979-8-4007-1454-2}
}

@inproceedings{ilharcoEditingModels2022,
  title = {Editing Models with Task Arithmetic},
  booktitle = {The {{Eleventh International Conference}} on {{Learning Representations}}},
  author = {Ilharco, Gabriel and Ribeiro, Marco Tulio and Wortsman, Mitchell and Schmidt, Ludwig and Hajishirzi, Hannaneh and Farhadi, Ali},
  year = 2022,
  month = sep,
  urldate = {2025-11-08},
  langid = {english}
}

@article{jiCriticalStudy2023,
  title = {A {{Critical Study}} on {{Data Leakage}} in {{Recommender System Offline Evaluation}}},
  author = {Ji, Yitong and Sun, Aixin and Zhang, Jie and Li, Chenliang},
  year = 2023,
  month = feb,
  journal = {ACM Trans. Inf. Syst.},
  volume = {41},
  number = {3},
  pages = {75:1--75:27},
  issn = {1046-8188},
  doi = {10.1145/3569930},
  urldate = {2025-03-05}
}

@inproceedings{kangSelfAttentiveSequential2018,
  title = {Self-{{Attentive Sequential Recommendation}}},
  booktitle = {2018 {{IEEE International Conference}} on {{Data Mining}} ({{ICDM}})},
  author = {Kang, Wang-Cheng and McAuley, Julian},
  year = 2018,
  month = nov,
  pages = {197--206},
  issn = {2374-8486},
  doi = {10.1109/ICDM.2018.00035},
  urldate = {2025-03-04}
}

@misc{kaplanScalingLaws2020,
  title = {Scaling {{Laws}} for {{Neural Language Models}}},
  author = {Kaplan, Jared and McCandlish, Sam and Henighan, Tom and Brown, Tom B. and Chess, Benjamin and Child, Rewon and Gray, Scott and Radford, Alec and Wu, Jeffrey and Amodei, Dario},
  year = 2020,
  month = jan,
  number = {arXiv:2001.08361},
  eprint = {2001.08361},
  primaryclass = {cs},
  publisher = {arXiv},
  doi = {10.48550/arXiv.2001.08361},
  urldate = {2025-11-14},
  archiveprefix = {arXiv}
}

@article{khanCrossDomain2017,
  title = {Cross {{Domain Recommender Systems}}: {{A Systematic Literature Review}}},
  shorttitle = {Cross {{Domain Recommender Systems}}},
  author = {Khan, Muhammad Murad and Ibrahim, Roliana and Ghani, Imran},
  year = 2017,
  month = jun,
  journal = {ACM Comput. Surv.},
  volume = {50},
  number = {3},
  pages = {36:1--36:34},
  issn = {0360-0300},
  doi = {10.1145/3073565},
  urldate = {2025-11-27}
}

@misc{kumarLLMPostTraining2025,
  title = {{{LLM Post-Training}}: {{A Deep Dive}} into {{Reasoning Large Language Models}}},
  shorttitle = {{{LLM Post-Training}}},
  author = {Kumar, Komal and Ashraf, Tajamul and Thawakar, Omkar and Anwer, Rao Muhammad and Cholakkal, Hisham and Shah, Mubarak and Yang, Ming-Hsuan and Torr, Phillip H. S. and Khan, Fahad Shahbaz and Khan, Salman},
  year = 2025,
  month = mar,
  number = {arXiv:2502.21321},
  eprint = {2502.21321},
  primaryclass = {cs},
  publisher = {arXiv},
  doi = {10.48550/arXiv.2502.21321},
  urldate = {2025-11-14},
  archiveprefix = {arXiv}
}

@inproceedings{linOrderagnosticIdentifier2025,
  title = {Order-Agnostic {{Identifier}} for {{Large Language Model-based Generative Recommendation}}},
  booktitle = {Proceedings of the 48th {{International ACM SIGIR Conference}} on {{Research}} and {{Development}} in {{Information Retrieval}}},
  author = {Lin, Xinyu and Shi, Haihan and Wang, Wenjie and Feng, Fuli and Wang, Qifan and Ng, See-Kiong and Chua, Tat-Seng},
  year = 2025,
  month = jul,
  series = {{{SIGIR}} '25},
  pages = {1923--1933},
  publisher = {Association for Computing Machinery},
  address = {New York, NY, USA},
  doi = {10.1145/3726302.3730053},
  urldate = {2025-12-23},
  isbn = {979-8-4007-1592-1}
}

@article{liSemanticConvergence2025,
  title = {Semantic {{Convergence}}: {{Harmonizing Recommender Systems}} via {{Two-Stage Alignment}} and {{Behavioral Semantic Tokenization}}},
  shorttitle = {Semantic {{Convergence}}},
  author = {Li, Guanghan and Zhang, Xun and Zhang, Yufei and Yin, Yifan and Yin, Guojun and Lin, Wei},
  year = 2025,
  month = apr,
  journal = {Proceedings of the AAAI Conference on Artificial Intelligence},
  volume = {39},
  number = {11},
  pages = {12040--12048},
  issn = {2374-3468},
  doi = {10.1609/aaai.v39i11.33311},
  urldate = {2025-11-08},
  copyright = {Copyright (c) 2025 Association for the Advancement of Artificial Intelligence},
  langid = {english}
}

@misc{liuDiagnosticGuidedDynamic2026,
  title = {Diagnostic-{{Guided Dynamic Profile Optimization}} for {{LLM-based User Simulators}} in {{Sequential Recommendation}}},
  author = {Liu, Hongyang and Sun, Zhu and Wei, Tianjun and Wang, Yan and Zhu, Jiajie and Qu, Xinghua},
  year = 2026,
  month = jan,
  number = {arXiv:2508.12645},
  eprint = {2508.12645},
  primaryclass = {cs},
  publisher = {arXiv},
  doi = {10.48550/arXiv.2508.12645},
  urldate = {2026-01-14},
  archiveprefix = {arXiv}
}

@inproceedings{liuLinearCombination2024,
  title = {Linear {{Combination}} of {{Saved Checkpoints Makes Consistency}} and {{Diffusion Models Better}}},
  booktitle = {The {{Thirteenth International Conference}} on {{Learning Representations}}},
  author = {Liu, Enshu and Zhu, Junyi and Lin, Zinan and Ning, Xuefei and Wang, Shuaiqi and Blaschko, Matthew B. and Yekhanin, Sergey and Yan, Shengen and Dai, Guohao and Yang, Huazhong and Wang, Yu},
  year = 2024,
  month = oct,
  urldate = {2025-12-03},
  langid = {english}
}

@inproceedings{rajputRecommenderSystems2023a,
  title = {Recommender {{Systems}} with {{Generative Retrieval}}},
  booktitle = {Thirty-Seventh {{Conference}} on {{Neural Information Processing Systems}}},
  author = {Rajput, Shashank and Mehta, Nikhil and Singh, Anima and Keshavan, Raghunandan Hulikal and Vu, Trung and Heldt, Lukasz and Hong, Lichan and Tay, Yi and Tran, Vinh Q. and Samost, Jonah and Kula, Maciej and Chi, Ed H. and Sathiamoorthy, Maheswaran},
  year = 2023,
  month = nov,
  urldate = {2025-11-08},
  langid = {english}
}

@article{sunLLM4RSRLarge2025,
  title = {{{LLM4RSR}}: {{Large Language Models}} as {{Data Correctors}} for {{Robust Sequential Recommendation}}},
  shorttitle = {{{LLM4RSR}}},
  author = {Sun, Yatong and Yang, Xiaochun and Sun, Zhu and Wang, Yan and Wang, Bin and Qu, Xinghua},
  year = 2025,
  month = apr,
  journal = {Proceedings of the AAAI Conference on Artificial Intelligence},
  volume = {39},
  number = {12},
  pages = {12604--12612},
  issn = {2374-3468},
  doi = {10.1609/aaai.v39i12.33374},
  urldate = {2026-01-15},
  copyright = {Copyright (c) 2025 Association for the Advancement of Artificial Intelligence},
  langid = {english}
}

@misc{tanPCRCAParallel2025,
  title = {{{PCR-CA}}: {{Parallel Codebook Representations}} with {{Contrastive Alignment}} for {{Multiple-Category App Recommendation}}},
  shorttitle = {{{PCR-CA}}},
  author = {Tan, Bin and Ge, Wangyao and Wang, Yidi and Liu, Xin and Burtoft, Jeff and Fan, Hao and Wang, Hui},
  year = 2025,
  month = sep,
  number = {arXiv:2508.18166},
  eprint = {2508.18166},
  primaryclass = {cs},
  publisher = {arXiv},
  doi = {10.48550/arXiv.2508.18166},
  urldate = {2025-11-08},
  archiveprefix = {arXiv}
}

@inproceedings{vandenoordNeuralDiscrete2017,
  title = {Neural Discrete Representation Learning},
  booktitle = {Proceedings of the 31st {{International Conference}} on {{Neural Information Processing Systems}}},
  author = {{van den Oord}, Aaron and Vinyals, Oriol and Kavukcuoglu, Koray},
  year = 2017,
  month = dec,
  series = {{{NIPS}}'17},
  pages = {6309--6318},
  publisher = {Curran Associates Inc.},
  address = {Red Hook, NY, USA},
  urldate = {2025-11-11},
  isbn = {978-1-5108-6096-4}
}

@inproceedings{wortsmanModelSoups2022,
  title = {Model Soups: Averaging Weights of Multiple Fine-Tuned Models Improves Accuracy without Increasing Inference Time},
  shorttitle = {Model Soups},
  booktitle = {Proceedings of the 39th {{International Conference}} on {{Machine Learning}}},
  author = {Wortsman, Mitchell and Ilharco, Gabriel and Gadre, Samir Ya and Roelofs, Rebecca and {Gontijo-Lopes}, Raphael and Morcos, Ari S. and Namkoong, Hongseok and Farhadi, Ali and Carmon, Yair and Kornblith, Simon and Schmidt, Ludwig},
  year = 2022,
  month = jun,
  pages = {23965--23998},
  publisher = {PMLR},
  issn = {2640-3498},
  urldate = {2025-11-08},
  langid = {english}
}

@inproceedings{yadavTIESMergingResolving2023,
  title = {{{TIES-Merging}}: {{Resolving Interference When Merging Models}}},
  shorttitle = {{{TIES-Merging}}},
  booktitle = {Thirty-Seventh {{Conference}} on {{Neural Information Processing Systems}}},
  author = {Yadav, Prateek and Tam, Derek and Choshen, Leshem and Raffel, Colin and Bansal, Mohit},
  year = 2023,
  month = nov,
  urldate = {2025-11-08},
  langid = {english}
}

@misc{yangModelMerging2024,
  title = {Model {{Merging}} in {{LLMs}}, {{MLLMs}}, and {{Beyond}}: {{Methods}}, {{Theories}}, {{Applications}} and {{Opportunities}}},
  shorttitle = {Model {{Merging}} in {{LLMs}}, {{MLLMs}}, and {{Beyond}}},
  author = {Yang, Enneng and Shen, Li and Guo, Guibing and Wang, Xingwei and Cao, Xiaochun and Zhang, Jie and Tao, Dacheng},
  year = 2024,
  month = sep,
  number = {arXiv:2408.07666},
  eprint = {2408.07666},
  primaryclass = {cs},
  publisher = {arXiv},
  doi = {10.48550/arXiv.2408.07666},
  urldate = {2025-11-13},
  archiveprefix = {arXiv}
}

@misc{yangQwen3Technical2025,
  title = {Qwen3 {{Technical Report}}},
  author = {Yang, An and Li, Anfeng and Yang, Baosong and Zhang, Beichen and Hui, Binyuan and Zheng, Bo and Yu, Bowen and Gao, Chang and Huang, Chengen and Lv, Chenxu and Zheng, Chujie and Liu, Dayiheng and Zhou, Fan and Huang, Fei and Hu, Feng and Ge, Hao and Wei, Haoran and Lin, Huan and Tang, Jialong and Yang, Jian and Tu, Jianhong and Zhang, Jianwei and Yang, Jianxin and Yang, Jiaxi and Zhou, Jing and Zhou, Jingren and Lin, Junyang and Dang, Kai and Bao, Keqin and Yang, Kexin and Yu, Le and Deng, Lianghao and Li, Mei and Xue, Mingfeng and Li, Mingze and Zhang, Pei and Wang, Peng and Zhu, Qin and Men, Rui and Gao, Ruize and Liu, Shixuan and Luo, Shuang and Li, Tianhao and Tang, Tianyi and Yin, Wenbiao and Ren, Xingzhang and Wang, Xinyu and Zhang, Xinyu and Ren, Xuancheng and Fan, Yang and Su, Yang and Zhang, Yichang and Zhang, Yinger and Wan, Yu and Liu, Yuqiong and Wang, Zekun and Cui, Zeyu and Zhang, Zhenru and Zhou, Zhipeng and Qiu, Zihan},
  year = 2025,
  month = may,
  number = {arXiv:2505.09388},
  eprint = {2505.09388},
  primaryclass = {cs},
  publisher = {arXiv},
  doi = {10.48550/arXiv.2505.09388},
  urldate = {2025-11-08},
  archiveprefix = {arXiv}
}

@misc{yangSparseMeets2025,
  title = {Sparse {{Meets Dense}}: {{Unified Generative Recommendations}} with {{Cascaded Sparse-Dense Representations}}},
  shorttitle = {Sparse {{Meets Dense}}},
  author = {Yang, Yuhao and Ji, Zhi and Li, Zhaopeng and Li, Yi and Mo, Zhonglin and Ding, Yue and Chen, Kai and Zhang, Zijian and Li, Jie and Li, Shuanglong and Liu, Lin},
  year = 2025,
  month = mar,
  number = {arXiv:2503.02453},
  eprint = {2503.02453},
  primaryclass = {cs},
  publisher = {arXiv},
  doi = {10.48550/arXiv.2503.02453},
  urldate = {2025-11-08},
  archiveprefix = {arXiv}
}

@article{yangUnifyingGenerative2025,
  title = {Unifying {{Generative}} and {{Dense Retrieval}} for {{Sequential Recommendation}}},
  author = {Yang, Liu and Paischer, Fabian and Hassani, Kaveh and Li, Jiacheng and Shao, Shuai and Li, Zhang Gabriel and He, Yun and Feng, Xue and Noorshams, Nima and Park, Sem and Long, Bo and Nowak, Robert D. and Gao, Xiaoli and Eghbalzadeh, Hamid},
  year = 2025,
  month = mar,
  journal = {Transactions on Machine Learning Research},
  issn = {2835-8856},
  urldate = {2026-01-13},
  langid = {english}
}

@inproceedings{yooContinualRecommender2025,
  title = {Continual {{Recommender Systems}}},
  booktitle = {Proceedings of the 34th {{ACM International Conference}} on {{Information}} and {{Knowledge Management}}},
  author = {Yoo, Hyunsik and Kang, Seongku and Tong, Hanghang},
  year = 2025,
  month = nov,
  series = {{{CIKM}} '25},
  pages = {6857--6860},
  publisher = {Association for Computing Machinery},
  address = {New York, NY, USA},
  doi = {10.1145/3746252.3761452},
  urldate = {2025-11-14},
  isbn = {979-8-4007-2040-6}
}

@inproceedings{yuLanguageModels2024,
  title = {Language {{Models}} Are {{Super Mario}}: {{Absorbing Abilities}} from {{Homologous Models}} as a {{Free Lunch}}},
  shorttitle = {Language {{Models}} Are {{Super Mario}}},
  booktitle = {Forty-First {{International Conference}} on {{Machine Learning}}},
  author = {Yu, Le and Yu, Bowen and Yu, Haiyang and Huang, Fei and Li, Yongbin},
  year = 2024,
  month = jun,
  urldate = {2025-11-08},
  langid = {english}
}

@article{zeghidourSoundStreamEndtoEnd2021,
  title = {{{SoundStream}}: {{An End-to-End Neural Audio Codec}}},
  shorttitle = {{{SoundStream}}},
  author = {Zeghidour, Neil and Luebs, Alejandro and Omran, Ahmed and Skoglund, Jan and Tagliasacchi, Marco},
  year = 2021,
  month = nov,
  journal = {IEEE/ACM Trans. Audio, Speech and Lang. Proc.},
  volume = {30},
  pages = {495--507},
  issn = {2329-9290},
  doi = {10.1109/TASLP.2021.3129994},
  urldate = {2025-11-27}
}

@inproceedings{zhaiActionsSpeak2024,
  title = {Actions {{Speak Louder}} than {{Words}}: {{Trillion-Parameter Sequential Transducers}} for {{Generative Recommendations}}},
  shorttitle = {Actions {{Speak Louder}} than {{Words}}},
  booktitle = {Proceedings of the 41st {{International Conference}} on {{Machine Learning}}},
  author = {Zhai, Jiaqi and Liao, Lucy and Liu, Xing and Wang, Yueming and Li, Rui and Cao, Xuan and Gao, Leon and Gong, Zhaojie and Gu, Fangda and He, Jiayuan and Lu, Yinghai and Shi, Yu},
  year = 2024,
  month = jul,
  pages = {58484--58509},
  publisher = {PMLR},
  issn = {2640-3498},
  urldate = {2025-04-19},
  langid = {english}
}

@inproceedings{zhangDualPhasePlaytimeguided2025,
  title = {Dual-{{Phase Playtime-guided Recommendation}}: {{Interest Intensity Exploration}} and {{Multimodal Random Walks}}},
  shorttitle = {Dual-{{Phase Playtime-guided Recommendation}}},
  booktitle = {Proceedings of the 33rd {{ACM International Conference}} on {{Multimedia}}},
  author = {Zhang, Jingmao and Zhao, Zhiting and Lin, Yunqi and Ma, Jianghong and Wei, Tianjun and Zhang, Haijun and Zhang, Xiaofeng},
  year = 2025,
  month = oct,
  series = {{{MM}} '25},
  pages = {6232--6241},
  publisher = {Association for Computing Machinery},
  address = {New York, NY, USA},
  doi = {10.1145/3746027.3755597},
  urldate = {2025-11-11},
  isbn = {979-8-4007-2035-2}
}

@misc{zhangQwen3Embedding2025,
  title = {Qwen3 {{Embedding}}: {{Advancing Text Embedding}} and {{Reranking Through Foundation Models}}},
  shorttitle = {Qwen3 {{Embedding}}},
  author = {Zhang, Yanzhao and Li, Mingxin and Long, Dingkun and Zhang, Xin and Lin, Huan and Yang, Baosong and Xie, Pengjun and Yang, An and Liu, Dayiheng and Lin, Junyang and Huang, Fei and Zhou, Jingren},
  year = 2025,
  month = jun,
  number = {arXiv:2506.05176},
  eprint = {2506.05176},
  primaryclass = {cs},
  publisher = {arXiv},
  doi = {10.48550/arXiv.2506.05176},
  urldate = {2025-11-27},
  archiveprefix = {arXiv}
}

@inproceedings{zhengAdaptingLarge2024,
  title = {Adapting {{Large Language Models}} by {{Integrating Collaborative Semantics}} for {{Recommendation}}},
  booktitle = {2024 {{IEEE}} 40th {{International Conference}} on {{Data Engineering}} ({{ICDE}})},
  author = {Zheng, Bowen and Hou, Yupeng and Lu, Hongyu and Chen, Yu and Zhao, Wayne Xin and Chen, Ming and Wen, Ji-Rong},
  year = 2024,
  month = may,
  pages = {1435--1448},
  issn = {2375-026X},
  doi = {10.1109/ICDE60146.2024.00118},
  urldate = {2025-11-08}
}

@misc{zhouOneRecV2Technical2025,
  title = {{{OneRec-V2 Technical Report}}},
  author = {Zhou, Guorui and Hu, Hengrui and Cheng, Hongtao and Wang, Huanjie and Deng, Jiaxin and Zhang, Jinghao and Cai, Kuo and Ren, Lejian and Ren, Lu and Yu, Liao and Zheng, Pengfei and Luo, Qiang and Wang, Qianqian and Hu, Qigen and Huang, Rui and Tang, Ruiming and Wang, Shiyao and Yang, Shujie and Wu, Tao and Li, Wuchao and Luo, Xinchen and Wang, Xingmei and Su, Yi and Wu, Yunfan and Cheng, Zexuan and Liu, Zhanyu and Zhang, Zixing and Zhang, Bin and Wang, Boxuan and Ma, Chaoyi and Song, Chengru and Wang, Chenhui and Chu, Chenglong and Wang, Di and Meng, Dongxue and Zang, Dunju and Yang, Fan and Zhang, Fangyu and Jiang, Feng and Zhang, Fuxing and Wang, Gang and Zhang, Guowang and Li, Han and Bao, Honghui and Cao, Hongyang and Huang, Jiaming and Chen, Jiapeng and Liu, Jiaqiang and Jia, Jinghui and Gai, Kun and Hu, Lantao and Zeng, Liang and Wang, Qiang and Zhou, Qidong and Zhang, Rongzhou and Wang, Shengzhe and He, Shihui and Yang, Shuang and Mao, Siyang and Huang, Sui and He, Tiantian and Gao, Tingting and Yuan, Wei and Liang, Xiao and Xu, Xiaoxiao and Liu, Xugang and Wang, Yan and Zhou, Yang and Wang, Yi and Liu, Yiwu and Song, Yue and Zhang, Yufei and Zhao, Yunfeng and Ling, Zhixin and Li, Ziming},
  year = 2025,
  month = oct,
  number = {arXiv:2508.20900},
  eprint = {2508.20900},
  primaryclass = {cs},
  publisher = {arXiv},
  doi = {10.48550/arXiv.2508.20900},
  urldate = {2025-11-14},
  archiveprefix = {arXiv}
}

@inproceedings{zhouRecBaseGenerative2025,
  title = {{{RecBase}}: {{Generative Foundation Model Pretraining}} for {{Zero-Shot Recommendation}}},
  shorttitle = {{{RecBase}}},
  booktitle = {Proceedings of the 2025 {{Conference}} on {{Empirical Methods}} in {{Natural Language Processing}}},
  author = {Zhou, Sashuai and Gan, Weinan and Liu, Qijiong and Lei, Ke and Zhu, Jieming and Huang, Hai and Xia, Yan and Tang, Ruiming and Dong, Zhenhua and Zhao, Zhou},
  editor = {Christodoulopoulos, Christos and Chakraborty, Tanmoy and Rose, Carolyn and Peng, Violet},
  year = 2025,
  month = nov,
  pages = {15598--15610},
  publisher = {Association for Computational Linguistics},
  address = {Suzhou, China},
  doi = {10.18653/v1/2025.emnlp-main.786},
  urldate = {2025-11-27},
  isbn = {979-8-89176-332-6}
}

@inproceedings{zhuangRobustlyOptimized2021,
  title = {A {{Robustly Optimized BERT Pre-training Approach}} with {{Post-training}}},
  booktitle = {Proceedings of the 20th {{Chinese National Conference}} on {{Computational Linguistics}}},
  author = {Zhuang, Liu and Wayne, Lin and Ya, Shi and Jun, Zhao},
  editor = {Li, Sheng and Sun, Maosong and Liu, Yang and Wu, Hua and Liu, Kang and Che, Wanxiang and He, Shizhu and Rao, Gaoqi},
  year = 2021,
  month = aug,
  pages = {1218--1227},
  publisher = {Chinese Information Processing Society of China},
  address = {Huhhot, China},
  urldate = {2025-11-14},
  langid = {english}
}
